\documentclass[9pt, oneside]{article}   	% use "amsart" instead of "article" for AMSLaTeX format
\usepackage{geometry}                		% See geometry.pdf to learn the layout options. There are lots.
\usepackage{graphicx}				% Use pdf, png, jpg, or eps§ with pdflatex; use eps in DVI mode
\usepackage{dsfont}								% TeX will automatically convert eps --> pdf in pdflatex		
\usepackage{amssymb}
\usepackage{algorithm}
\usepackage[noend]{algpseudocode}
\usepackage{hyperref}
\usepackage{todonotes}
\usepackage{pf2}

\title{State-Based $\infty$P-Set Conflict-Free Replicated Data Type}
\author{Erick Lavoie}
\date{April, 4th, 2023}							% Activate to display a given date or no 

\begin{document}
\maketitle

\begin{abstract}
The 2P-Set Conflict-Free Replicated Data Type (CRDT) supports two phases for each possible element: in the first phase an element can be added to the set and the subsequent additions are ignored; in the second phase an element can be removed after which it will stay removed forever regardless of subsequent additions and removals. We generalize the 2P-Set to support an infinite sequence of alternating additions and removals of the same element. In the presence of concurrent additions and removals on different replicas, all replicas will eventually converge to the longest sequence of alternating additions and removals that follows causal history. 

The idea of converging on the longest-causal sequence of opposite operations had already been suggested in the context of an undo-redo framework but the design was neither given a name nor fully developed. In this paper, we present the full design directly, using nothing more than the basic formulation of state-based CRDTs. We also show the connection between the set-based definition of 2P-Set and the counter-based definition of the $\infty$P-Set with simple reasoning. We then give detailed proofs of convergence. The underlying \textit{grow-only dictionary of grow-only counters} on which the $\infty$P-Set is built may be used to build other state-based CRDTs. In addition, this paper should be useful as a pedagogical example for designing state-based CRDTs, and might help raise the profile of CRDTs based on \textit{longest sequence wins}.
\end{abstract}

\section{Introduction}

Conflict-Free Replicated Data Types (CRDTs)~\cite{shapiro:hal-00932836} are replicated mutable objects that are designed to ensure converge to the same state \textit{eventually}, \textit{i.e.}, at some point in the future after updates have stopped, and \textit{automatically}, \textit{i.e.} using deterministic conflict-resolution rules in the presence of concurrent updates. 

The 2P-Set~\cite{shapiro:inria-00555588} is a replicated set that supports two phases for each possible element: in the first phase, an element can be added to the set and any subsequent addition of the same element is going to be ignored; in the second phase, an element that is already in the set can be removed and any subsequent additions and removals of the same element are going to be ignored.

The C-Set~\cite{aslan:inria-00594590} enables addition of an element after removal but has counter-intuitive behaviour for some concurrent updates~\cite{bieniusa:hal-00769554}: two replicas may independently issue sequences of additions and removals both ending in an addition but eventually converge to a state in which the element is actually not in the set.

The Observe-Remove-Set (OR-Set)~\cite{shapiro:inria-00555588} also enables addition of an element after removal but requires tracking each addition operation with a unique identifier, resulting in memory usage proportional to the number of concurrent additions of the same element.

The Last-Writer-Wins-Set (LWW-Set)~\cite{shapiro:inria-00555588} also enables addition of an element after removal but requires two sets, ordering of concurrent operations according to a timestamp, and garbage collection of stale state tuples.

The T-Set~\cite{Dolan2020undoable} also enables addition of an element after removal and represents whether an element is in a set or not by associating it with 0 or 1 with modifications performed as modulo 2 operations. However, its operation-based formulation relies on causal delivery to determine which operation happened last. A similar state-based design, based on the mapping of the state of elements in the set to integers, had previously been  suggested in the context of a framework for undo-redo operations on state-based CRDTs~\cite{Yu2019undo}. However, the complete design had not been presented and no name had been given to it.

In this paper, we present the complete design of a state-based set design that supports an infinite sequence of additions and removals, which we call the $\infty$P-Set due to its straight-forward extension of the 2P-Set semantics. An $\infty$P-Set represents whether an element is in a set as positive integers within a \textit{grow-only dictionary of grow-only counters}: if the associated counter is odd the element is in the set, otherwise if no counter is present or the counter is even, the element is not in the set.

Similar to the T-Set~\cite{Dolan2020undoable} and unnamed predecessor~(\cite{Yu2019undo}, Section 5.1), our design derives from the following two observations. First, given operations \textit{add(e)} and \textit{remove(e)} on a set $S$ implemented as a CRDT, the implementation of $S$ has only two possible states: one in which an element $e$ is in the set $S$ ($e \in S$) and one in which an element $e$ is not in the set $S$ ($e \notin S$). When $e \in S$, only a \textit{remove(e)} modifies the state, and when $e \notin S$, only an \textit{add(e)} modifies the state. When only observing state changes, any arbitrary sequence of \textit{add(e)} and \textit{remove(e)} is actually equivalent to a strict alternating sequence of \textit{add(e)} and \textit{remove(e)}. Second, two replicas performing the same sequence of \textit{add(e)} and \textit{remove(e)} will reach the same state, even without communication. When two replicas have sequences of different lengths, we can see the smaller as a prefix of the longer and ignore the smaller, therefore \textit{the longest sequence eventually wins} and all replicas converge to the final state of the longest sequence. Note that the longest sequence need not happen on a single replica, it may actually have been generated through a sequence of replicas according to the causal history. Note also that the longest sequence will be given priority even if a concurrent but more recent shorter sequence exists, \textit{e.g.}, according to a global timestamp.  

In the rest of this paper\footnote{Sources for this paper are also available here: \url{https://github.com/cn-uofbasel/infinite-P-Set}. Pull-requests to suggest corrections are welcome.}, we make the following contributions:
\begin{itemize}
	\item We present the complete design of an $\infty$P-Set. In contrast to the aforementioned undo-redo framework~\cite{Yu2019undo} which introduces its own formalism, we rely on nothing more than the base formalism of state-based CRDTs (Section~\ref{sec:specification}). Our explanation of the connection between the 2P-Set and the $\infty$P-Set is also simpler than the presentation of the T-Set~\cite{Dolan2020undoable}: the latter relies on group theory which most undergrad computer science and practitioners are unlikely to have been exposed to;
	\item We provide a proof that $\infty$P-Set is indeed a state-based CRDT, with a high-level overview in Section~\ref{sec:proofs:convergence} and a detailed presentation in Appendix~\ref{sec:longer-proofs:convergence}. The underlying \textit{grow-only dictionary of grow-only counters} may also be useful for building more sophisticated state-based CRDTs. The proofs are written in the structured style suggested by Lamport~\cite{lamport2012write} and only assume background on discrete mathematics, making them useful as pedagogical example for undergrad students and practitioners;
\end{itemize}

In addition, we also discuss the lower memory consumption of an $\infty$P-Set compared to an OR-Set or LWW-Set (Section~\ref{sec:resource-consumption}). We then conclude with a summary and some directions for future work (Section~\ref{sec:conclusion}).

\section{$\infty$P-Set}
\label{sec:specification}

The $\infty$P-Set enables each possible element to be either in or out of the set, following a possibly infinite sequence of addition and removal operations. Concurrent additions and removals are resolved by having the longest sequence of alternating additions and removals win. 

In this section, we first present how the design is an extension of a 2P-Set (Section~\ref{sec:intuition}), the state and operations specifying its behaviour (Section~\ref{sec:data-type}), its causal and concurrent behaviour (Section~\ref{sec:causal-concurrent-behaviour}), and the system model under which it is valid (Section~\ref{sec:system-model}).

\subsection{From 2P-Set to $\infty$P-Set}
\label{sec:intuition}

Our design is a generalization of the behaviour of the 2P-Set~\cite{shapiro:inria-00555588}, which we show as follows. 

In a 2P-Set, an element $e$ can either be \textit{in} a set $S$ ($e \in S$) or \textit{out} of the same set ($e \notin S$). An element is initially out of the set. After an addition \textit{add(e)}, the element is in the set, which corresponds to the first phase. Any subsequent addition of $e$ does not change the state of $S$. After a removal \textit{remove(e)}, $e$ is then considered out of the set forever, which corresponds to the second phase.

A 2P-Set is implemented by combining 2 \textit{grow-only sets} $A$ and $R$. $A$ tracks additions, and is accordingly called the \textit{add-set}; $R$ tracks removals, and is accordingly called the \textit{remove-set} (or \textit{tombstones} set). The \textit{add(e)} operation adds $e$ to the add-set $A$. The \textit{remove(e)} operation adds $e$ to the remove-set $R$, only if $e$ is already in the add-set $A$. The element $e$ is considered in $S$ if and only if $e$ is in the difference of $A$ and $R$, \textit{i.e.} $e \in S \Leftrightarrow e \in (A\backslash R)$. Merging two states $(A_1, R_1)$ and $(A_2, R_2)$ is simply the union of their components, \textit{i.e.} $(A_1 \cup A_2, R_1 \cup R_2)$.

The 2P-Set can be extended to a 4P-Set by adding another pair of \textit{grow-only sets}, so that the state of $S$ can be described by two pairs of add-remove-sets $(A_1,R_1,A_2,R_2)$.  This enables $e$ to be added, then removed, then added again, then removed one last time before staying removed forever. During an \textit{add(e)} operation, if $e$ is not in any of the components, it is added to $A_1$; if it is in both in $A_1$ and $R_1$, it is added to $A_2$; otherwise, the \textit{add(e)} is ignored. Similarly, during a \textit{remove(e)} operation, if $e$ is only in $A_1$, $e$ is added to $R_1$; if $e$ is in $A_1$, $R_1$, and $A_2$, $e$ is added to $R_2$; otherwise the \textit{remove(e)} is ignored. The element $e$ is considered in the set $S$ if and only if the set difference of either pair includes $e$, \textit{i.e.} $e \in S \Leftrightarrow e \in  ((A_1 \backslash R_1) \cup (A_2 \backslash R_2))$. Merging is simply the element-wise union of components.

The 2P-Set can be extended to an $\infty$P-Set by adding an infinite number of add-remove-set pairs $(A_1, R_1, A_2, R_2, \dots)$. Manipulating an infinite number of sets is unwieldy, so instead we can track the number of add-sets and remove-sets 
$e$ is included in: if $e$ is in all add-sets up to $A_i$ and all remove-sets up to $R_{j}$, we simply store the state of $e$ as $(i, j)$. Furthermore, because the state of $e$ is strictly alternating between being in and out of $S$, and $i$ and $j$ monotonically grow as $e$ is further added and removed, we can simply map the state of $e$ to a single positive integer $k=i+j$. The element $e$ is in $S$ if and only if $k$ is odd otherwise it is out of $S$, \textit{i.e.} $e \in S \Leftrightarrow odd(k)$. Merging the state of $e$ in $S$ and $e$ in $S'$ corresponds to taking the maximum value of the corresponding counters, \textit{i.e.} $\textit{max}(k, k')$, which accordingly is determined by the longest sequence of \textit{add(e)} and \textit{remove(e)} that happened across replicas. 

\subsection{State and Operations}
\label{sec:data-type}

The behaviour of the $\infty$P-Set is listed in Algorithm~\ref{alg:inf-p-set} following existing conventions~\cite{shapiro:inria-00555588}.  The state of the CRDT is implemented as a dictionary $D$ that 
maps unique elements to a single positive integer. The corresponding set $S$ is obtained by querying the dictionary, \textit{i.e.} $S=\texttt{query}(D)$. The same set $S$ might be implemented by different valid dictionaries so the inverse relationship is ambiguous. All operations are specified in a functional style. When an operation modifies the state, its first argument is the current state $D$ and its return value is the modified state $D'$ or $D''$. This makes the algorithm easier to associate with the proofs (Section~\ref{sec:proofs}). A practical implementation might instead encapsulate the state, in an object for example.

Operations are the followings:
\begin{itemize}

	\item $D=\texttt{Initialize}()$ creates a new $\infty$P-Set replica. After initialization, the state of the replica, $D$, is an empty dictionary. 

	\item $S=\texttt{Query}(D)$ returns the set $S$ corresponding to the dictionary $D$. $S$ contains the elements that are currently in $S$, \textit{i.e.}, all elements in $D$ associated with an \textit{odd} integer counter.

	\item $D'=\texttt{Add}(D,e)$ adds the element $e$ to the set represented by $D$, returning a new dictionary $D'$.  Adding $e$ when $e$ is already in the set, \textit{i.e.}, $D[e]$ exists and is odd, is ignored and $D'=D$. Otherwise, if $e$ had never been added before ($e$ is not in the keys of $D$): $e$ is added and its counter is initialized to 1, \textit{i.e.}, $D'[e] = 1$. Finally, if $e$ has been added before but the last operation was a \texttt{remove} ($e$ is in $D$ and $D[e]$ is even): $D'[e] = D[e] + 1$, making it odd and effectively adding $e$ back in the set.

\item $D'=\texttt{Remove}(D, e)$ removes the element $e$ from the set represented by $D$, returning a new dictionary $D'$. The element $e$ might not be in the set either because it was never added before, in which case $e$ won't be in $D$, or because it was previously removed, in which case $D[e]$ is even. In both cases, the remove operation is ignored. Otherwise, the last operation on $e$ was an \texttt{add} and $D[e]$ is odd. In that case, $D'[e]=D[e]+1$, making $D'[e]$ even and effectively removing $e$ from the set.

\item $b=\texttt{Compare}(D, D')$ returns $b=\texttt{true}$ if $D'$ includes all operations that were performed on $D$ and possibly more. Otherwise, it returns $b=\texttt{false}$. The comparison is true if and only if the keys of $D$ are a subset of those of $D'$ and all the counters associated to every element of $D$ are smaller or equal than corresponding counters in $D'$. If $\texttt{compare}(D,D')= \texttt{compare}(D',D)$ there are two possibilities: if both are \texttt{true} then both dictionaries are actually equal and have received the same sequences of operations; otherwise, the ordering is not defined which means that $D$ and $D'$ have been modified concurrently but not merged yet.

\item $D''=\texttt{Merge}(D, D')$ combines $D$ and $D'$ such that $D''$ will be greater than both but as small as possible, effectively incorporating all operations that happened to both $D$ and $D'$. If an element $e$ is  in $D'$ but not in $D$, $e$ is added with the associated counter from $D'$, and inversely. Otherwise, the counter of $e$ is chosen as the maximum value of the corresponding counters in both $D$ and $D'$, which corresponds to the longest sequence of alternating \textit{add(e)} and \textit{remove(e)} applied on either $D$ or $D'$.
\end{itemize}

\begin{algorithm}
\begin{algorithmic}[1]
   \Function{initialize}{}
    	\State $D \leftarrow \{ \}$ \Comment{Dictionary mapping each element $e$ to an integer counter $c$}
    	\State \Return $D$
    \EndFunction
    \State
    \Function{query}{D}
        \State $S \leftarrow \{~ e \textbf{~for all~} e \in \textit{keys}(D) ~\textbf{if}~ D[e]$~is odd~ $\}$ 
        \State \Return $S$
    \EndFunction
    \State
    \Function{add}{D, e}
        \State $D' \leftarrow \textit{copy}(D)$
    	\If{$e \notin \textit{keys}(D)$}
		\State $D'[e] \leftarrow 1$
	\ElsIf{$D[e]$ is even}
		\State $D'[e] \leftarrow D[e] + 1$
	\EndIf
	\State \Return $D'$
    \EndFunction
    \State
    \Function{remove}{D, e}
        \State $D' \leftarrow \textit{copy}(D)$
	\If{$e \in \textit{keys}(D)$ ~\textbf{and}~ $D[e]$ is odd}
		\State $D'[e] \leftarrow D[e] + 1$
	\EndIf
	\State \Return $D'$
    \EndFunction
    \State
    \Function{compare}{$D$, $D'$} \Comment{true iff $D \leq D'$}
    	\State \Return $\textit{keys}(D) \subseteq \textit{keys}(D') \wedge \bigwedge_{e \in \textit{keys}(D)} D[e] \leq D'[e]$
    \EndFunction
    \State
    \Function{merge}{$D$, $D'$}
        \State $D'' \leftarrow \{ \}$
        \State $K'' \leftarrow  \textit{keys}(D) \cup \textit{keys}(D')$
       
	\For{$e \in K''$}
		\If{$e \in \textit{keys}(D) \wedge e \in \textit{keys}(D')$}
			\State $D''[e] \leftarrow \textit{max}(D[e], D'[e])$
		\ElsIf{$e \in \textit{keys}(D)$}
			\State $D''[e] \leftarrow D[e]$
		\Else
			\State $D''[e] \leftarrow D'[e]$
		\EndIf
	\EndFor
	\State \Return $D''$	
    \EndFunction
\end{algorithmic}
\caption{\label{alg:inf-p-set} $\infty$P-Set CRDT (State-based)}
\end{algorithm}

\subsection{Causal and Concurrent Behaviour}
\label{sec:causal-concurrent-behaviour}

The full behaviour of an $\infty$P-Set implies the followings. Assume an initial state $D$ and two operations $\textit{add}_r(e)$ and $\textit{remove}_{r'}(e)$ respectively happening on replicas $r$ and $r'$ and a final state $D'$ after both replicas $r$ and $r'$ have merged. The corresponding sets are $S=\texttt{query}(D)$ and $S'=\texttt{query}(D')$:
\begin{enumerate}
	\item If $\textit{add}_r(e)$ happens before $\textit{remove}_{r'}(e)$ ($\textit{add}_r(e)\rightarrow \textit{remove}_{r'}(e)$) then eventually $e \notin S'$ on $r$ and $r'$ regardless of the state of $e$ in $S$;
	\item  If $\textit{remove}_{r'}(e)$ happens before $\textit{add}_{r}(e)$ ($\textit{remove}_{r'}(e) \rightarrow  \textit{add}_r(e)$) then eventually $e \in S'$ on $r$ and $r'$ regardless of the state of $e$ in $S$;
	\item else both are concurrent ($add_r(e) ~||~ remove_{r'}(e)$) and the result depends on $S$:
	    \begin{enumerate}
		\item	 If $e \in S$ ($D[e]$ is odd) then $add_r(e)$ has no effect, $remove_{r'}(e)$ wins, and eventually $e \notin S'$ on $r$ and $r'$;
		\item Else $e \notin S$ (either $e \notin \textit{keys}(D)$ or $D[e]$ is even), $remove_{r'}(e)$  has no effect, $add_r(e)$ wins, and eventually $e \in S'$ on $r$ and $r'$.
	    \end{enumerate}
\end{enumerate}

If replicas $r$ and $r'$ are in different states, there still exists a common state $D$ they shared in the past (the initial state if nothing else), and from that point, the longest alternating sequence $D \rightarrow \textit{add(e)} \rightarrow \textit{remove(e)}\rightarrow \textit{add}(e) \rightarrow \dots $ across replicas determines the final state.

\subsection{System Model}
\label{sec:system-model}

As customary for state-based CRDTs~\cite{shapiro:inria-00555588}, Algorithm~\ref{alg:inf-p-set} only assumes an underlying unreliable communication channel that eventually delivers  a message (possibly mutiple times) if the message is sent infinitely often. This message is used to merge the state of two different replicas. In addition, transitive connectivity between replicas is assumed so that any update may reach any replica, even if indirectly. Finally, replicas may become unresponsive for arbitrarily long and infinitely often as long as they do eventually recover and merge state updates after each failure.

Note that Algorithm~\ref{alg:inf-p-set} does not tolerate arbitrary faults: a malicious replica may single-handedly determine the state of an element in the set by simply choosing a counter value arbitrarily large.

\section{Proof Sketches}
\label{sec:proofs}

In this section, we provide proof sketches. Please see Appendix~\ref{sec:longer-proofs:convergence} for more detailed proof steps.

\subsection{Convergence}
\label{sec:proofs:convergence}

To establish convergence, we need to show that a state-based CRDT definition is a \textit{monotonic semi-lattice}~\cite{shapiro:hal-00932836}. A state-based CRDT definition combines the followings:
\begin{itemize}
	\item The \textit{set of possible states} $\mathds{S}$: in the case of an $\infty$P-Set, this is the set of all possible dictionaries that associate an element $e$, out of all possible elements $E$ that could be stored, to an integer greater than zero that represents the state of $e$ in the set $S = \texttt{query}(D)$. The state of a given replica $D$ at any time is always one of those of $\mathds{S}$, \textit{i.e.} $D \in \mathds{S}$;
	\item A \textit{partial order} $\leq$ with which we can compare two states $D, D' \in \mathds{S}$ such that $D \leq D'$ if and only if the the set of updates that led to $D$ is a subset of the updates that led to $D'$. This is equivalent to say that $D \leq D'$ if and only if $D$ is the same as $D'$ or has happened before $D'$;
	\item An \textit{initial state} $D_0$ for all replicas such that $D_0 \in \mathds{S}$ onto which all possible sequences of updates are applied. In the case of a $\infty$P-Set, $D_0$ is an empty dictionary and created by \texttt{Initialize};
	\item A \textit{set of query operations} $Q$ that do not modify the state: which in the case of an $\infty$P-Set is $\{ \texttt{Query}, \texttt{Compare} \}$;
	\item A \textit{set of update operations} $U$ that do modify the state (but not necessarily for all combination of state and arguments): which in the case of an $\infty$P-Set is $\{ \texttt{Add}, \texttt{Remove} \}$;
	\item A \textit{merge} operation $m$ that may potentially modify the state: which in the case of an $\infty$P-Set is ${ \texttt{Merge} }$.
\end{itemize}

To establish that our state-based CRDT definition is a monotonic semi-lattice, we need to verify three propositions: First, that all possible states are organized in a semi-lattice $\mathds{L}$ ordered by less-or-equal relationship implemented by \texttt{compare}. This is a pre-requisite for the next two properties. Second, that merging any two states $D$ and $D'$ computes the \textit{Least Upper Bound} (LUB) of $D$ and $D'$ in $\mathds{L}$. This ensures that the merge is \textit{commutative}, \textit{associative}, and \textit{idempotent}, providing \textit{safety}, \textit{i.e.} that replicas will agree on the final state regardless of ordering, delays, or duplication of merge operations. Third, that all operations modify the state $D$ of a replica such that the new state $D'$ is either equal or larger than the previous state $D$ in $\mathds{L}$ (\textit{monotonicity}). This ensures all state changes will be eventually reflected in the new state of all replicas, either because the same update(s) will have concurrently been applied or because the new state will be the result of a merge. Assuming an underlying communication medium that ensures new states to be eventually delivered to other replicas, the three propositions combined ensure both \textit{liveness} and \textit{safety}: all state changes are going to be replicated on all replicas \textit{and} all replicas will agree on the final state automatically, i.e. \textit{strong eventual consistency}~\cite{shapiro:hal-00932836}.

%\textbf{Theorem}: \textit{The $\infty$P-Set (Alg.~\ref{alg:inf-p-set}) is a state-based (convergent) conflict-free replicated data type.}

\begin{proof}
        \prove{The $\infty$P-Set (Alg.~\ref{alg:inf-p-set}) is a state-based (convergent) CRDT.}
	\pfsketch ~The $\infty$P-Set is the composition of state-based grow-only sets for dictionary keys and max-counters for dictionary values. The three properties of \textit{ordering}, \textit{least upper bound}, and \textit{monotonicity} that are sufficient to define a state-based CRDT are the conjunction of corresponding properties on grow-only sets and max-counters.
	
	\label{proof:state-crdt}
	\step{label-ordering}{\textit{(Ordering)} Ordering $\mathds{S}$ by \texttt{compare} forms a semi-lattice $\mathds{L}$.}
	\begin{proof}
		\pfsketch~ \texttt{Compare} is the conjunction of the $\subseteq$ and $\leq$ relationships, respectively forming partial orders on sets and natural numbers that compose the possible states. The conjunction of partial orders is also a partial order.		
	\end{proof}

	\step{label-lub}{\textit{(Least-Upper Bound)} \texttt{Merge} of states $D$ and $D'$ computes the LUB of $D$ and $D'$ in $\mathds{L}$.}
	\begin{proof}
			\pfsketch~ \texttt{Merge} is the conjunction of performing the union on keys and maximum values on counters, which both compute the least upper bound in $\mathds{L}$, respectively on keys and counters. In conditions in which the maximum value cannot be computed, the corresponding part of the state is equal to one of the state input, \textit{e.g.} $D$, and the relationship to the other state input, \textit{e.g.} $D'$, can safely be ignored.
	\end{proof}

	\step{label-monotonicity}{\textit{(Monotonicity)} All operations that may generate a new state, when applied on state $D$ and any possible arguments, result in a new state either equal or larger than $D$ in $\mathds{L}$ according to \texttt{compare}.}
	\begin{proof}
		\pfsketch~ Every operation either does not have an input state as argument, does not generate a new state, or generate a new state that is equal or larger than all state inputs.
	\end{proof}

	\qedstep
	\begin{proof}
		\pf By definition, \stepref{label-ordering}, \stepref{label-lub}, \stepref{label-monotonicity} imply that $\infty$P-Set is a state-based CRDT. 
	\end{proof}
\end{proof}

\subsection{Known Anomalies Are Avoided}

We show here that anomalies previously identified~\cite{bieniusa:hal-00769554} are correctly handled by $\infty$P-Set. 

First, the specific execution in which the effect of a \textit{remove} is ignored even if it happened after an \textit{add} from a different replica (Fig. 1(a) in~\cite{bieniusa:hal-00769554}) does not happen with $\infty$P-Set because as mentioned in Section~\ref{sec:causal-concurrent-behaviour}, in the presence of a causal relationship, the remove will apply.

Second,  the $\infty$P-Set does not exhibit the counter-intuitive merging behaviour of C-Sets~\cite{bieniusa:hal-00769554} in which two replicas may both issue a sequence of \textit{add(e)} and \textit{remove(e)} that ends in an \textit{add(e)} while $e$ is not in the set after synchronization (Fig. 1(b) in~\cite{bieniusa:hal-00769554}):

\begin{proof}
	\prove{Merging any two $\infty$P-Sets with states $D$ and $D'$, whose last operation was an \textit{add(e)} (respectively \textit{remove(e)}), always results in a state $D''$ such that $e \in \texttt{query}(D'')$ (respectively $e \notin \texttt{query}(D'')$).}

	\pf~ When observing the corresponding dictionaries $D$, $D'$, and $D''$ is which both $D[e]$ and $D'[e]$ are odd because their last operation was an \textit{add(e)}, there are two cases:
	\begin{pfenum}
    		\item $D[e]=D'[e]$ and after a merge $D''[e]=D[e]=D'[e]$;
    		\item $D[e]>D'[e]$ or $D[e]<D[e']$: after a merge $D''[e]=max(D[e], D''[e])$
	\end{pfenum}   

	In both cases $D''[e]$ is also odd and $e \in  \texttt{query}(D'')$.  The proof for \textit{remove(e)} being the last operation of both sequences is similar, except that $D[e]$ and $D'[e]$ are even.

\end{proof}

Third, whenever two replicas $r$ and $r'$ issue the same sequence of \textit{add(e)} and \textit{remove(e)} starting from the same state $D$ (Fig. 1 b and c in~\cite{bieniusa:hal-00769554}), $r$ and $r'$ will both end in the same state $D'$ after synchronization, regardless at which point the synchronization happens and even if no synchronization happens, because the same sequence will result in the same counter for $e$ on all replicas.

\section{Resource Consumption}
\label{sec:resource-consumption}

The overhead of the $\infty$P-Set is a single integer per element plus the associated containers. It can be implemented as a set of tuples $(e,c)$ where $e$ is the element and $c$ the associated counter. It can also be implemented as a hash-map: current language implementations usually represent integer literals as tagged pointers allowing counts to reach counts up to 
$\frac{\textit{register size}}{2^\textit{tag bit-length}}$. In practice this reaches over billions on today's architectures which should cover most practical sequences of additions and removals. Up to the limit of literal integer representations, the memory consumption per element is constant.

In contrast to Observe-Remove Sets (OR-Sets)~\cite{shapiro:inria-00555588}, each add operation does not need to be tracked with a separate unique identifier and stored in a set. The OR-set uses memory proportional to the number of additions that have happened since the last removal, with one unique identifier for each, in addition to the size of the set container for each element that has ever been added in the set $S$.

In contrast to Last-Writer-Wins Sets (LWW-Sets)~\cite{shapiro:inria-00555588}, no timestamps are required and only a single set (or dictionary) is needed, and garbage collection of stale tuples is unnecessary.

\section{Conclusion and Future Work}
\label{sec:conclusion}

We have presented the state-based $\infty$P-Set CRDT. This approach requires only a single extra integer per element to track whether the element was last added or removed. We have shown that it converges and that it avoids anomalies that affected some other designs, such as the C-Set. We have also shown that it uses less memory than OR-Sets or LWW-Sets. We have also provided proofs that are more accessible than previous papers and established a \textit{grow-only dictionary of grow-only counters} as a basic block for other state-based CRDTs. We plan to extend the design to arbitrary (Byzantine) failures in the future and apply the design in applications to validate whether the semantics of \textit{longest sequence wins} suit both programmers and end users.

\section{Acknowledgements}
\label{sec:acknowledgements}

We thank Christian F. Tschudin for fostering a research environment allowing detours and playfulness in the process, as well as providing financial support for this work and feedback on early drafts. 

The current specification using a single counter and state testing on even-oddness was jointly rediscovered in collaboration with Christian F. Tschudin and Ramon Locher during a CRDT Seminar at University of Basel during the Spring Semester of 2023. It was a small disappointment, after a literature review, to realize we were 4 years too late to claim originality.

We would also like to thank Jannick Heisch, Fabrizio Parrillo, and Osman Biçer for feedback on early versions of the paper, and the material I had prepared for the CRDT seminar. Those discussions have informed the design and presentation.

\bibliographystyle{plain}
\bibliography{main}

\newpage
\appendix

\section{Detailed Proofs}

\subsection{Convergence}
\label{sec:longer-proofs:convergence}

\begin{proof}
\define{\begin{pfenum}
		\item \textit{(elements}) $E$ is the set of possible elements that can be stored in an $\infty$P-Set.
		\item \textit{(natural numbers)} $\mathds{N}$ are the natural numbers, \textit{i.e.} the set of integers greater or equal to $0$.
		\item \label{dict-state} \textit{(possible states)} The state of an $\infty$P-Set is a set $D$ of tuples $(e,c)$ such that $e \in E$ and counter value $c \in \mathds{N}$. $D$ represents a dictionary. The state space of all possible dictionaries, $\mathds{S}$, is the power set of the cartesian product of $E$ and $\mathds{N}$ ($\mathds{S} = \mathcal{P}(E \times \mathds{N})$) with the constraint that for every $D \in \mathds{S}$ and among all tuples $(e,c) \in D$, a given element $e$ appears at most once.
		\item \label{dict-keys} \textit{(keys)} The keys of $D$ is the set of elements in $D$, \textit{i.e.} $keys(D) = \{ e : (e,c) \in D \}$.
		\item \label{dict-counter} \textit{(counter)} The counter associated to $e$ in $D$, written $D[e]$, is the counter $c$ corresponding to the tuple $(e,c) \in D$, \textit{i.e.} $D[e] = c \Leftrightarrow (e,c) \in D$.
		\item \label{dict-assignment} \textit{(dictionary assignment)} Assigning a counter value $c$ to a dictionary key $e$, written $D[e] \leftarrow c$, performs one of two actions: 
		\begin{pfenum}
			\item If $(e,c') \notin \textit{keys}(D)$, it adds a new tuple $(e,c)$ in $D$, \textit{i.e.} $D=D \cup \{ (e,c) \}$;
			\item otherwise, it replaces the tuple with $(e,c)$, \textit{i.e.} $D=D \backslash \{ (e, c') \} \cup \{ (e, c) \}$.
		\end{pfenum}
		\item \textit{(possible keys)} Dictionary keys $\mathds{K} = \{ \textit{keys}(D) : D \in \mathds{S} \}$.
		\item \textit{(possible counters}) Counter values $\mathds{C} = \{ D[e] :  D \in \mathds{S} \wedge e \in \textit{keys}(D) \}$.
	\end{pfenum}}
\end{proof}

\begin{proof}
        \prove{The $\infty$P-Set (Alg.~\ref{alg:inf-p-set}) is a state-based (convergent) CRDT.}
	\pfsketch ~The $\infty$P-Set is the composition of state-based grow-only sets for dictionary keys and max-counters for dictionary values. The three properties of \textit{ordering}, \textit{least upper bound}, and \textit{monotonicity} that are sufficient to define a state-based CRDT are the conjunction of corresponding properties on grow-only sets and max-counters.

	\label{proof:state-crdt}
\end{proof}

\subsubsection{Ordering}
\label{sec:longer-proofs:ordering}

\begin{proof}
	\assume{\begin{pfenum}
		\item $D \in \mathds{S}$ and $D' \in \mathds{S}$ in the arguments of \texttt{compare}.
	\end{pfenum}}
	\prove{Ordering $\mathds{S}$ by \texttt{compare} forms a semi-lattice $\mathds{L}$.}
		\pfsketch~ \texttt{Compare} is the conjunction of the $\subseteq$ and $\leq$ relationships, respectively forming partial orders on sets and natural numbers that compose the possible states. The conjunction of partial orders is also a partial order. A semi-lattice is a partial order on a set.

		%\begin{proof}
		%	\pf The state of any dictionary $D \in \mathds{S}$ can be viewed as a set of tuples $(e,c)$ composed of an element $e \in E$ and a natural counter $c$ such that $c \geq 0$.
		%\end{proof}
		\step{label-1}{All possible dictionary keys $\mathds{K}$ are partially ordered by $\subseteq$.}
		\begin{proof}
		
			\step{label-1.1}{$\mathds{K}$ is a powerset of elements, \textit{i.e.} $\mathds{K} = \mathcal{P}(E)$}
			\begin{proof}
				By definitions \ref{dict-state} and \ref{dict-keys}.
			\end{proof}
			
			\step{label-1.2}{Sets are partially ordered by $\subseteq$.}
			\begin{proof}
				The subset relationship $\subseteq$ is \textit{reflexive}, \textit{transitive}, and \textit{antisymmetric},  therefore sets ordered by $\subseteq$ are partially-ordered.
			\end{proof}
			
			\qedstep
			 \begin{proof}
			  	By \stepref{label-1.1} and \stepref{label-1.2}.
			\end{proof}
		\end{proof}
		
		\step{label-2}{All possible counter values $\mathds{C}$ are partially ordered by $\leq$.}
		\begin{proof}
		
			\step{label-2.1}{$\mathds{C}$ is the set of natural numbers, \textit{i.e.} $\mathds{C} = \mathds{N}$}
			\begin{proof}
				By definitions \ref{dict-state} and \ref{dict-counter}.
			\end{proof}
			
			\step{label-2.2}{Natural numbers are partially ordered by $\leq$.}
			\begin{proof}
				The less or equal relationship $\leq$ is \textit{reflexive}, \textit{transitive}, and \textit{antisymmetric}.
			\end{proof}

			\qedstep
			\begin{proof}
				By \stepref{label-2.1} and \stepref{label-2.2}.
			\end{proof}
		\end{proof}
		
		\step{label-3}{
			\assume{\begin{pfenum}
				\item Relations $\alpha$ and $\beta$ are partial orders over set $S$.
				\item $a, b \in S$.
			\end{pfenum}}
			\define{\begin{pfenum}
				\item \label{label-def-conj-partial-orders} Conjunction of relations $\alpha$ and $\beta$, written $\tau$, as $a ~\tau~ b = (a ~\alpha~ b) \wedge (a ~\beta~ b)$
			\end{pfenum}}
			\prove{$\tau$ is a partial order over $S$.}
		}
		\begin{proof}
			\pfsketch~ $\tau$ is also \textit{reflexive}, \textit{transitive}, and \textit{antisymmetric} because of the associativity of $\wedge$. The three properties are sufficient to define a partial order.
			\step{label-3.0}{$\alpha$ and $\beta$ are both \textit{reflexive}, \textit{transitive}, and \textit{antisymmetric}.}
			\begin{proof}
				By definition because partial order relations must be \textit{reflexive}, \textit{transitive}, and \textit{antisymmetric}.
			\end{proof}
	
			\step{label-3.1}{$\tau$ is reflexive: $a ~\tau~ a = true$.}
			\begin{proof}
				$a ~\tau~ a = (a ~\alpha~ a) \wedge (a ~\beta~ a) = true$ because  $\alpha$ and $\beta$ are also reflexive.
			\end{proof}
			
			\step{label-3.2}{$\tau$ is transitive:  $(a ~\tau~ b) \wedge (b ~\tau~ c) \Rightarrow a ~\tau~ c$.}
			\begin{proof}
				\step{label-3.2.1}{$(a ~\tau~ b) \wedge (b ~\tau~ c) =  ((a ~\alpha~ b) \wedge (a ~\beta~ b)) \wedge ((b ~\alpha~ c) \wedge (b ~\beta~ c))$}
				\begin{proof}
					By substitution of $\tau$ definition.
				\end{proof}
				\step{label-3.2.2}{$ ((a ~\alpha~ b) \wedge (a ~\beta~ b)) \wedge ((b ~\alpha~ c) \wedge (b ~\beta~ c)) = ((a ~\alpha~ b) \wedge (b ~\alpha~ c)) \wedge ((a ~\beta~ b) \wedge (b ~\beta~ c))$}
				\begin{proof}
					By associativity of $\wedge$ (and) relation.
				\end{proof}
				\step{label-3.2.3}{$((a ~\alpha~ b) \wedge (b ~\alpha~ c)) \wedge ((a ~\beta~ b) \wedge (b ~\beta~ c)) = (a ~\alpha~ c) \wedge (a ~\beta~ c)$}
				\begin{proof}
					By transitivity of $\alpha$ and $\beta$ (\stepref{label-3.0}).
				\end{proof}
				
				\step{label-3.2.4}{$ (a ~\alpha~ c) \wedge (a ~\beta~ c) = (a ~\tau~ c)$}
				\begin{proof}
					By definition \ref{label-def-conj-partial-orders} in \stepref{label-3}.
				\end{proof}
				
				\qedstep
				\begin{proof}
					\stepref{label-3.2.1} $=$ \stepref{label-3.2.2} $=$ \stepref{label-3.2.3} $=$ \stepref{label-3.2.4}, therefore $(a ~\tau~ b) \wedge (b ~\tau~ c) \Rightarrow  (a ~\tau~ c)$
				\end{proof}
			\end{proof}
			
			\step{label-3.3}{$\tau$ is antisymmetric: $(a ~\tau~ b) \wedge (b ~\tau~ a) \Rightarrow (a = b)$.}
			\begin{proof}
				\step{label-3.3.1}{$(a ~\tau~ b) \wedge (b ~\tau~ a) =  ((a ~\alpha~ b) \wedge (a ~\beta~ b)) \wedge ((b ~\alpha~ a) \wedge (b ~\beta~ a))$}
				\begin{proof}
					By substitution of $\tau$ definition.
				\end{proof}
				
				\step{label-3.3.2}{$ ((a ~\alpha~ b) \wedge (a ~\beta~ b)) \wedge ((b ~\alpha~ a) \wedge (b ~\beta~ a)) = ((a ~\alpha~ b) \wedge (b ~\alpha~ a)) \wedge ((a ~\beta~ b) \wedge (b ~\beta~ a))$}
				\begin{proof}
					By associativity of $\wedge$ (and) relation.
				\end{proof}
				
				\step{label-3.3.3}{$((a ~\alpha~ b) \wedge (b ~\alpha~ a)) \wedge ((a ~\beta~ b) \wedge (b ~\beta~ a)) = (a = b) \wedge (a = b)$}
				\begin{proof}
					By antisymmetry of $\alpha$ and $\beta$ (\stepref{label-3.0}).
				\end{proof}
				
				\step{label-3.3.4}{$ (a = b) \wedge (a = b) = (a = b)$}
				\begin{proof}
					Tautologie.
				\end{proof}
				
				\qedstep
				\begin{proof}
					\stepref{label-3.3.1} $=$ \stepref{label-3.3.2} $=$ \stepref{label-3.3.3} $=$ \stepref{label-3.3.4}
				\end{proof}
			\end{proof}
			
			\qedstep
			\begin{proof}
				By \stepref{label-3.1}, \stepref{label-3.2}, and \stepref{label-3.3}, which is the definition of a partial order.
			\end{proof}
		\end{proof}
			
			\step{label-4}{
				\assume{\begin{pfenum}
					\item \label{subset} $\textit{keys}(D) \subseteq \textit{keys}(D')$
				\end{pfenum}}
				\prove{$\bigwedge_{e \in \textit{keys}(D)} D[e] \leq D'[e]$ is a partial order}}
			\begin{proof}
				Because of Assumption~\ref{subset}, $e \in \textit{keys}(D')$ as well, therefore both $D[e]$ and $D'[e]$ are defined. Because \stepref{label-2}, $\leq$ is a partial order over counters of $D$ and $D'$ for element $e$. Because \stepref{label-3}, the pairwise conjunction of partial orders is also a partial order, and therefore the conjunction over all keys of $D$ is also a partial order. Note that contrary to the next step, there is no logical dependency between the partial orderings of individual keys so the conjunction directly applies.
			\end{proof}
			
			\step{label-5}{The conjunction of $\textit{keys}(D) \subseteq \textit{keys}(D')$ and \stepref{label-4} is a partial order.}
			\begin{proof}
				\pfsketch~ There is a logical dependency from the left-hand side of the conjunction to the right-hand side because the left-hand side is a necessary assumption to compute the right-hand side. The partial order is therefore the conjunction of both when the left-hand side is true, otherwise only the left-hand side is used.
				\step{label-5.1}{\case{$\textit{keys}(D) \subseteq \textit{keys}(D')$ is true}}
				\begin{proof}
					Then Assumption~\ref{subset} of \stepref{label-4} is satisfied and the resulting partial order is their conjunction. $\textit{keys}(D) \subseteq \textit{keys}(D')$ is a partial order because \stepref{label-1}. The conjunction with \stepref{label-4} is also a partial order because of \stepref{label-3}.
				\end{proof}
				
				\step{label-5.1}{\case{$\textit{keys}(D) \subseteq \textit{keys}(D')$ is false}}
				\begin{proof}
					The conjunction will always be false, so the additional partial order induced by \stepref{label-4} is irrelevant. Therefore, the partial order is only defined by $\textit{keys}(D) \subseteq \textit{keys}(D')$. Note that the right-hand side must be ignored because Assumption~\ref{subset} is not satisfied and the result is therefore not defined.
				\end{proof}
				
				\qedstep
				\begin{proof}
					All possible cases of the conjunction define partial orders, so the conjunction of $\textit{keys}(D) \subseteq \textit{keys}(D')$ and \stepref{label-4} is a partial order.
				\end{proof}
			\end{proof}
			
			\qedstep
			\begin{proof}
				\texttt{compare} is a partial order over $\mathds{S}$ because its definition is completely covered by \stepref{label-5}. A semi-lattice is the combination of possible states and a partial order, therefore both conditions are sufficient to define $\mathds{L}$.
			\end{proof}			
	\end{proof}

\subsubsection{Least Upper Bound (LUB)}
\label{sec:longer-proofs:lub}

\begin{proof}
	\assume{\begin{pfenum}
		\item $D \in \mathds{S}$ and $D' \in \mathds{S}$
	\end{pfenum}}	
	\prove{
		\texttt{Merge} of $D$ and $D'$ computes their LUB $D''$ in $\mathds{L}$.
	}
	\begin{proof}
		\pfsketch~ \texttt{Merge} is the conjunction of performing the union on keys and maximum values on counters, which both compute the least upper bound, respectively on keys and values. In conditions when the maximum value cannot be computed, the corresponding part of the state is equal to one of the state input, \textit{e.g.} $D$, and the relationship to the other state input, \textit{e.g.} $D'$, can safely be ignored.
		\step{label-1}{
		\assume{\begin{pfenum}
			\item $\mathcal{P}(E)$ is the power set of elements in $E$.
			\item $S \in \mathcal{P}(E)$ and $S' \in \mathcal{P}(E)$.
			\item $\subseteq$ is a partial order on $\mathcal{P}(E)$ forming a semi-lattice $\mathds{L}_{\mathcal{P}(E)}$.
		\end{pfenum}}
		\prove{$S'' = S \cup S'$ is the least upper bound of $S$ and $S'$ in $\mathds{L}_{\mathcal{P}(E)}$.}}
		\begin{proof}
			\step{label-1.1}{$S \subseteq S''$}
			\begin{proof}
				Because $S \subseteq (S \cup S') = S''$ (by cases on $S \subset S', S=S', S \supset S'$).
			\end{proof}
			
			\step{label-1.2}{$S' \subseteq S''$}
			\begin{proof}
				Because $S' \subseteq (S \cup S') = S''$ (by cases on $S \subset S', S=S', S \supset S'$).
			\end{proof}
			
			\step{label-1.3}{$\nexists S''' \in \mathcal{P}(E)$ such that $S''' \subset S'' \wedge S \subseteq S''' \wedge S' \subseteq S''' $}
			\begin{proof}
				By contradiction: Let's assume a contrario that there exists such a $S'''$. There must exist $e \in S''$ such that $e \notin S'''$ because $S''' \subset S''$ and therefore smaller. Since $S'' = S \cup S'$, either $e \in S$, $e \in S'$ or both. In that case, it is not possible for $S'''$ to be a superset of both $S$ and $S'$ since it is missing $e$. $S''$ must therefore be the smallest superset of both $S$ and $S'$ in $\mathds{L}_{\mathcal{P}(E)}$.
			\end{proof}

			\qedstep
			\begin{proof}
				The conjunction of \stepref{label-1.1}, \stepref{label-1.2}, and \stepref{label-1.3} is the definition of a least upper bound.	
			\end{proof}
		\end{proof}
		
		\step{label-2}{
		\assume{\begin{pfenum}
			\item $x \in \mathds{N}$ and $x' \in \mathds{N}$.
			\item $\leq$ is a total order (and partial as well) on $\mathds{N}$ forming a semi-lattice $\mathds{L}_\mathds{N}$.
		\end{pfenum}}
		\prove{$x'' = \textit{max}(x,x')$ is the least upper bound of $x$ and $x'$ in $\mathds{L}_\mathds{N}$.}}
\begin{proof}
			\step{label-2.1}{$x \leq x''$}
			\begin{proof}
				Because $x \leq \textit{max}(x,x') = x''$ (by cases on $x<x', x=x', x>x'$).
			\end{proof}
			
			\step{label-2.2}{$x' \leq  x''$}
			\begin{proof}
				Because $x' \leq \textit{max}(x,x') = x''$ (by cases on $x<x', x=x', x>x'$).
			\end{proof}
			
			\step{label-2.3}{$\nexists x''' \in \mathds{N}$ such that $x''' < x'' \wedge x \leq x''' \wedge x' \leq x''' $}
			\begin{proof}
				By contradiction: Let's assume a contrario that there exists such a $x'''$. Because $x''' < x''$, $x'''$ must be smaller than $x$, $x'$, or both because $x'' = \textit{max}(x,x')$. However, this contradicts $ x \leq x'''$,  $x' \leq x'''$ or both. Therefore, $x''$ must be the smallest $\mathds{N}$ that is also greater than both $x$ and $x'$ in $\mathds{L}_\mathds{N}$.
			\end{proof}

			\qedstep
			\begin{proof}
				The conjunction of \stepref{label-2.1}, \stepref{label-2.2}, and \stepref{label-2.3} is the definition of a least upper bound.	
			\end{proof}
		\end{proof}

		\step{label-3}{
		$\textit{keys}(D'')$ is the least upper bound of $\textit{keys}(D)$ and $\textit{keys}(D')$.}
		\begin{proof}
			$\textit{keys}(D), \textit{keys}(D'), \textit{keys}(D'')$ are sets; $\textit{keys}(D'') = \textit{keys}(D) \cup \textit{keys}(D')$, and \stepref{label-1}, together imply that $\textit{keys}(D'')$ is the least upper bound of $\textit{keys}(D)$ and $\textit{keys}(D')$ in $\mathds{L}_{\mathcal{P}(E)}$.
		\end{proof}
		
		\step{label-3}{
		For every $e \in \textit{keys}(D'')$ such that $e \in \textit{keys}(D) \wedge e \in \textit{keys}(D')$, $D''[e]$ is the least upper bound of $D[e]$ and $D'[e]$.}
		\begin{proof}
			$D[e], D'[e], D''[e]$ are natural numbers, $D''[e] = \textit{max}(D[e], D'[e])$, and \stepref{label-2}.
		\end{proof}
		
		\step{label-4}{
		\texttt{compare}$(D,D'')$ is true ($D \leq D''$, according to $\mathds{L}$) }
		\begin{proof}
			\step{label-4.1}{$\textit{keys}(D) \subseteq \textit{keys}(D'')$}
			\begin{proof}
				$K'' = \textit{keys}(D'') = (\textit{keys}(D) \cup \textit{keys}(D')) \supseteq \textit{keys}(D)$
			\end{proof}
		
			\step{label-4.2}{$\forall e \in \textit{keys}(D) : D[e] \leq D''[e]$}
			\begin{proof}
				\step{label-4.2.1}{$\forall e \in \textit{keys}(D) : e \in \textit{keys}(D'')$}
				\begin{proof}
					Because \stepref{label-4.1}. Necessary condition for following cases.
				\end{proof}
				
				\step{}{\case{$D''[e] = \textit{max}(D[e], D'[e])$}}
				\begin{proof}
					$\textit{max}(D[e], D'[e]) \geq D[e]$ implies $D[e] \leq D''[e]$.
				\end{proof}
				
				\step{}{\case{$D''[e] = D[e]$}}
				\begin{proof}
					Trivially: $D''[e] = D[e]$ implies $D[e] \leq D''[e]$.
				\end{proof}
				
				\qedstep
				\begin{proof}
				 	Note that $D''[e] = D'[e]$, which happens when $e \in \textit{keys}(D'') \wedge e \in \textit{keys}(D') \wedge e \notin \textit{keys}(D)$ does not matter for \texttt{compare}$(D, D'')$ because $D[e] \leq D''[e]$ is only tested for $e \in \textit{keys}(D)$.
				\end{proof}
			\end{proof}
			
			\qedstep
			\begin{proof}
				By the conjunction of \stepref{label-4.1} and \stepref{label-4.2}.
			\end{proof}
		\end{proof}
		
		\step{label-5}{
		\texttt{compare}$(D',D'')$ is true ($D' \leq D''$, according to $\mathds{L}$) }
		\begin{proof}
			Same argument as \stepref{label-4} but on $D'$ instead of $D$.
		\end{proof}
		
		\step{label-6}{
		$\nexists D''' \in \mathds{S}$ such that $D''' < D'' \wedge D \leq D''' \wedge D' \leq D'''$ in $\mathds{L}$.}
		\begin{proof}
			\step{label-6.1}{$\nexists \textit{keys}(D''')$ such that $\textit{keys}(D''') \subset \textit{keys}(D'') \wedge \textit{keys}(D) \subseteq \textit{keys}(D''') \wedge \textit{keys}(D') \subseteq \textit{keys}(D''') $.}
			\begin{proof}
				$\textit{keys}(D'') = \textit{keys}(D) \cup \textit{keys}(D')$ and \stepref{label-1} imply that $\textit{keys}(D'')$ is a least upper bound, which by definition implies there is no $D'''$ satisfying the above conditions.
			\end{proof}
			
			\step{label-6.2}{$\nexists e \in \textit{keys}(D'') \wedge c \in \mathds{N} : c < D''[e] \wedge D[e] \leq c \wedge D'[e] \leq c$}
			\begin{proof}
				\pfsketch~ There are three possible cases for $e \in \textit{keys}(D'')$ and none allow $c \in \mathds{N}$ with the required conditions.
				\step{}{\case{$e \in \textit{keys}(D) \wedge e \in \textit{keys}(D')$}}
				\begin{proof}
					$D''[e] = \textit{max}(D[e], D'[e])$, by \stepref{label-2} implies $D''[e]$ is a least upper bound, which by definition implies that such a $c$ does not exists.
				\end{proof}
				
				\step{}{\case{$e \notin \textit{keys}(D) \wedge e \in \textit{keys}(D')$}}
				\begin{proof}
					$D''[e] = D'[e]$ therefore a $c \in \mathds{N}$ cannot be both smaller than $D'']e]$ and larger than $D'[e]$. Therefore, $D[e] \leq c$ (which is undefined) can be safely ignored because $c < D''[e] \wedge D'[e] \leq c$ is always false.
				\end{proof}
				
				\step{}{\case{$e \in \textit{keys}(D) \wedge e \notin \textit{keys}(D')$}}
				\begin{proof}
					$D''[e] = D[e]$ therefore a $c \in \mathds{N}$ cannot be both smaller than $D'']e]$ and larger than $D[e]$. Therefore,  $D'[e] \leq c$ (which is undefined) can be safely ignored because $c < D''[e] \wedge D[e] \leq c$ is always false.

				\end{proof}
				
				\qedstep
				\begin{proof}
					The case $e \notin \textit{keys}(D) \wedge e \notin \textit{keys}(D')$ contradicts $e \in \textit{keys}(D'')$ since $\textit{keys}(D'') = \textit{keys}(D) \cup \textit{keys}(D')$. Since $\nexists$ is a negation and the following conditions in \stepref{label-6.2} are always false for all cases, \stepref{label-6.2} is always true.
				\end{proof}
			\end{proof}
			
			\qedstep
			\begin{proof}
				Because \texttt{compare} is the conjonction of conditions on keys and counters, respectively covered by \stepref{label-6.1} and \stepref{label-6.2} there cannot be such a $D'''$. Therefore $D''$ is the least upper bound of $D$ and $D'$ in $\mathds{L}$.
			\end{proof}
		\end{proof}
		
		\qedstep
		\begin{proof}
			The conjunction of \stepref{label-4}, \stepref{label-5}, and \stepref{label-6} is the definition of a least upper bound, therefore  \texttt{merge} computes the least upper bound of $D$ and $D'$ in $\mathds{L}$.
		\end{proof}	
	\end{proof}

\end{proof}

\subsubsection{Monotonicity}
\label{sec:longer-proofs:monotonicity}

\begin{proof}
	\assume{\begin{pfenum}
		\item $D$ is the current state	
	\end{pfenum}}	
	\prove{All operations that may generate a new state, when applied on state $D$ and any possible arguments, result in a new state either equal or larger than $D$ in $\mathds{L}$ according to \texttt{compare}.}

		\pfsketch~ By case, because every operation either does not have an input state as argument, does not generate a new state, or generate a new state that is equal or larger than all state inputs.
		\step{}{\case{$D = \texttt{Initialize}()$}}
		\begin{proof}
			Initializes a new $D$, but not from an existing state, so monotonicity does not apply.
		\end{proof}
		
		\step{}{\case{$S = \texttt{Query}(D)$}}
		\begin{proof}
			\texttt{query} does not generate a new state.
		\end{proof}
		
		\step{}{\case{$D' = \texttt{Add}(D,e) \geq D$}}
		\begin{proof}
			\step{}{\case{$e \notin \textit{keys}(D)$}}
			\begin{proof}
				The key $e$ is added to $D$ with counter $c=1$, therefore $D' > D$ because $D'$ keys are a superset of $D$'s keys.
			\end{proof}
			
			\step{}{\case{$e \in \textit{keys}(D) \wedge D[e]$ is even}}
			\begin{proof}
				The counter $D[e]$ is increased by 1, therefore $D' > D$.
			\end{proof}
			
			\step{}{\case{$e \in \textit{keys}(D) \wedge D[e]$ is odd}}
			\begin{proof}
				No keys or counters are modified, therefore $D=D'$.
			\end{proof}
			
			\qedstep
			\begin{proof}
				Covers all possible cases: $e$ in or not in the keys, and counter $D[e]$ is odd or even if $e$ present.
			\end{proof}
		\end{proof}
		
		\step{}{\case{$D' = \texttt{Remove}(D,e) \geq D$}}
		\begin{proof}
			\step{}{\case{$e \notin \textit{keys}(D)$}}
			\begin{proof}
				No keys or counters are modified, therefore $D=D'$.
			\end{proof}
			
			\step{}{\case{$e \in \textit{keys}(D) \wedge D[e]$ is odd}}
			\begin{proof}
				The counter $D[e]$ is increased by 1, therefore $D' > D$.
			\end{proof}
			
			\step{}{\case{$e \in \textit{keys}(D) \wedge D[e]$ is even}}
			\begin{proof}
				No keys or counters are modified, therefore $D=D'$.
			\end{proof}
			
			\qedstep
			\begin{proof}
				Covers all possible cases: $e$ in or not in the keys, and counter $D[e]$ is odd or even if $e$ present.
			\end{proof}
		\end{proof}
		
		\step{}{\case{$b = \texttt{Compare}(D,D')$}}
		\begin{proof}
			\texttt{compare} does not generate a new state.
		\end{proof}
		
		\step{}{\case{$D'' = \texttt{Merge}(D,D') : D'' \geq D \wedge D'' \geq D'$}}
		\begin{proof}
			By definition, because \texttt{merge} computes the least upper bound (Appendix~\ref{sec:longer-proofs:lub}).
		\end{proof}
	
	\qedstep
	\begin{proof}
		All functions of Algorithm~\ref{alg:inf-p-set} have been covered.
	\end{proof}	
\end{proof}

\end{document}